\documentclass[twocolumn]{aastex62}
\graphicspath{{./}{}}

\received{\today}
\revised{\today}
\accepted{\today}
\submitjournal{ApJ}

\shorttitle{Gaia-ASAS-SN Cepheids}
\shortauthors{Inno et al.}

\begin{document}

\title{The Gaia-ASAS-SN classical Cepheids sample: I. Sample Selection}
\correspondingauthor{Laura Inno}
\email{laura.inno@uniparthenope.it}

\author[0000-0002-0786-7307]{Laura Inno}
\affil{Science and Technology Department, Parthenope University of Naples, Naples, Italy}
\altaffiliation{INAF-Osservatorio Astronomico di Capodimonte,\\ 
Salita Moraliello, Napoli}

\author{Hans-Walter Rix}
\affil{Max-Planck-Institut f\"ur Astronomie,
69117, Heidelberg, Germany}
 \author{K. Z. Stanek}
\affil{Department of Astronomy, The Ohio State University, 140 West 18th Avenue, Columbus, OH 43210, USA}
\altaffiliation{Center for Cosmology and Astroparticle Physics, \\ The Ohio State University, 191 W. Woodruff Avenue, \\Columbus, OH 43210, USA}
 \author{T. Jayasinghe}
\affil{Department of Astronomy, The Ohio State University, 140 West 18th Avenue, Columbus, OH 43210, USA}
\altaffiliation{Center for Cosmology and Astroparticle Physics, \\ The Ohio State University, 191 W. Woodruff Avenue,\\ Columbus, OH 43210, USA}
\author{E. Poggio}
\affil{Osservatorio Astrofisico di Torino, Istituto Nazionale di Astrofisica (INAF), I-10025 Pino Torinese, Italy}
\altaffiliation{Universit\'e C\^ote d’Azur, \\ Observatoire de la C\^ote d’Azur, \\ CNRS, Laboratoire Lagrange, France}
 \author{R. Drimmel}
\affil{Osservatorio Astrofisico di Torino, Istituto Nazionale di Astrofisica (INAF), I-10025 Pino Torinese, Italy}
\author{A. Rotundi}
\affil{Science and Technology Department, Parthenope University of Naples, Naples, Italy}

\begin{abstract}
We present a well-defined and characterized all-sky sample of classical Cepheids in the Milky Way, 
obtained by combining two time-domain all-sky surveys: Gaia DR2 \citep{gaiacoll18} and ASAS-SN \citep{shappee14}.
We first use  parallax and variability information from Gaia to select $\sim$ 30,000 bright (G$<$17) Cepheid candidates with $M_K<-1$. 
We then analyze their ASAS-SN V-band lightcurves, 
determining periods, and classifying the lightcurves using their Fourier parameters.
This results in $\sim$1900 likely Galactic Cepheids, which we estimate to be $\gtrsim 90$\% complete and pure within our adopted selection criteria. This is the largest all-sky sample of Milky Way Cepheids that has such a well-characterized selection function, needed for population modeling and for systematic spectroscopic follow-up foreseen with SDSS-V.
About 130 of these Cepheids have not been documented in the literature even as possible candidates.

\end{abstract}


\section{Introduction} 
\label{sec:intro}

Classical Cepheids are fundamental calibrators of the Cosmic distance scale. They are also near-ideal tracers of the ``young'' Milky Way ($<300$~Myrs), as  luminosities, distances, masses and ages can be inferred from their lightcurves. Despite this fundamental importance, classical Cepheids had been poorly mapped in our Galaxy until very recently, 
with only $\sim$600 identified over the past decades 
\citep[e.g][and references therein]{laney92,berdnikov04,jayasinghe18,matsunaga18}.
The advent of automated all-sky, time-domain surveys over
the past years have brought about a quite dramatic change: 
new catalogs based on surveys such as ATLAS, ASAS-SN, WISE and ZTF have been released,
resulting in the identification of about 500 \citep{heinze18,udalski19}, 200 \citep{jayasinghe18},1300 \citep{chen19}  and 429\citep{chen20} new potential classical Cepheids, respectively.
Smaller, dedicated surveys can offer higher sample completeness and purity, especially in highly extincted regions at low Galactic latitude: for example,
the near-IR VISTA survey {\it Via Lactea} reported the discovery of 
$\sim$200 Cepheids beyond the Galactic Center \citep{dekany15a,dekany15b, dekany19}; 
and the IRSF-SIRIUS survey identified 25 obscured Cepheids in the inner Galaxy \citep{matsunaga15,tanioka17,inno19}.

Gaia DR2 \citep{gaiacoll18} led to a catalog of variable stars with time-series photometry, which
includes $\sim$1,200 classical Cepheids within a Galactic latitude of $\pm$20 degree \citep{clementini18}. 
Recently, \citet[][hereinafter R19]{ripepi18} has presented a new classification of all variable sources in the Gaia DR2 catalog, selecting a cleaner sample of $\sim$680 {\sl Fundamental Mode} (FU) and 147  {\sl First Overtone Mode} (1OV) Cepheids. Among those, $\sim$250 had not been reported earlier. 
\citet[][hereinafter U18]{udalski19} and \citet{sos20} recently released the OGLE-IV Collection of Variable Stars (OGLE-IV CVS), 
which includes about 1,800 newly identified classical Cepheids.
On the basis of this sample, and additional photometric 
data from the ATLAS and ASAS-SN surveys, \citet[][hereinafter S19b]{skowron19a,skowron19b} 
were able to accurately map the structure of the young Galactic disk 
in 3D for the first time: they determined some structural parameters of the Galactic disk, such as
the scale height, the flaring and the warp, between 
a range of Galactocentric radii between 3 and 20 kpc.

One aspect in common to all these catalogs is that they do not provide a quantitative characterization of their selection function: in which directions and to what distance do Cepheids have a quantifiable probability of entering the sample, as a function of their luminosity, with its implied mass and age, their variability amplitude, effective temperature (or color) and light-curve shape, given also the presence of possible dust reddening. Yet, this is indispensable basis for any inference about the Cepheid's spatial and age distribution and also their number density in the Galaxy.

In this work we set out to identify a sample of classical Galactic Cepheids that is a) all-sky, b) extensive ($\sim$2000 objects), and c) has a well-defined, consistent and reproducible selection function.
We do this by combining data from two available all-sky optical time-domain surveys, Gaia and ASAS-SN. Specifically, we lay out a clear algorithmic method to identify an initial set of Cepheid candidates through a query of the Gaia DR2 main catalog that combines {\it rms} variability,  {\it color}-variability, luminosity and intrinsic color. We then analyse ASAS-SN time-resolved photometry \citep{shappee14}, to identify and classify the likely classical Cepheids through a Fourier analysis of their lightcurves. We consciously did not choose the \emph{largest possible classical Cepheid catalog} as our goal, and consciously left out known Cepheids if they did not fulfill our selection criteria. This catalog will be used by SDSS-V for systematic, all-sky follow-up of Galactic disk Cepheids.

The method is described in Section~\ref{sec:catalog_new},
while the characterization of the output Catalog in terms of completeness and purity,
and its comparison to literature catalogs, is provided in 
Section~\ref{sec:catalog_comp_1}, 
Section~\ref{sec:catalog_purity_1} and \ref{sec:catalog_comp} respectively.

\section{The selection of Galactic Classical Cepheids from Gaia DR2 data and ASAS-SN time series} 
\label{sec:catalog_new}

Our goal for this paper is to identify an extensive, large all-sky sample of Galactic classical Cepheids with a well-characterized selection function, and modest or small contamination. This is needed for systematic multi-epoch spectroscopic follow-up, which has just been initiated with SDSS-V \citep{Kollmeier2017}.

We do this by developing a selection function for application to Gaia DR2 $\times$ ASAS-SN data, which encapsulates most of the basic properties that could plausibly be used to identify Cepheids:
\begin{enumerate}
    \item they are luminous, with absolute $K_{\rm{S}}$-band magnitude of $M_K\lesssim -1$
    \item their optical lightcurves vary periodically, with periods of days to about two months, and with range of (optical, or G-band) peak-to-peak amplitudes around a characteristic value of 0.5 mag.
    \item they are pulsating variables, where luminosity changes go along with changes in the effective photosphere temperatures: they vary in color, as they vary in magnitude.
    \item they are young ($\tau \lesssim 250$~Myrs), and should move on disk-like orbits. 
    \item their lightcurve shapes depend on their period and on the nature of their dominant pulsation mode (fundamental mode, or $1^{st}$ overtone), which can be compactly characterized by a Fourier expansion \citep[see e.g.][]{simon1981,antonello1987,mantegazza1992,deb09,inno15}
\end{enumerate}

We implement this by first applying a subset of these criteria to the Gaia DR2 data, resulting in an extensive, but highly contaminated, candidate list, and then analyzing the candidate's ASAS-SN lightcurves for identification and classification.

\subsection{Initial Candidate Selection from Gaia DR2}
\label{sec:DR2_selection}

The set of Cepheid \emph{candidates} across the entire sky, can be obtained straightforwardly by a single query of the Gaia data archive\footnote{\url{gea.esac.esa.int/archive/}}.
We now describe the rationale behind each element of this query.

The selection function should reflect that Cepheids are luminous, for which we adopt $M_K^{lim}=-1$. But for many candidates the Gaia parallaxes will be ``marginal''. Therefore we require that $M_K^{lim} < -1$, even if the true parallax were 1$\sigma$ larger than the best estimate. This leads to:
\begin{equation}
    \varpi + \sigma\varpi < 10^{0.2 (10. - \tilde{m}_K - M_K^{lim})},
\end{equation}
where $\tilde{m}_K\equiv$~{\tt phot\_g\_mean\_mag} -  1.72({\tt bp\_rp}) + 0.07({\tt bp\_rp)$^2$} is an empirically determined prediction of $m_K$, based on Gaia colors alone; this approximation appears good to $0.1$~mag even once optical extinction values reach several magnitudes. This effects a reddening-insensitive estimate of the apparent and ultimately absolute magnitude, yet avoids involving a second catalog (e.g. 2MASS) in the selection.

While Gaia DR2 is generally not a time-domain data release, limited information about variability is encoded in the data products. Specifically, we use $\Delta_{G}$ 
\begin{equation}
   \Delta_{G}\equiv \sqrt{phot\_g\_n\_obs} \times \frac{phot\_g\_mean\_flux\_error}{phot\_g\_mean\_flux} 
   \label{eqn:variability_definition}
\end{equation}
as proxy for {\it rms} variability in the Gaia $G$-band. This variability measure was proposed by
\citet{deason17} to identify RR Lyrae. One can also construct analogous quantities $\Delta_{BP}$ and $\Delta_{RP}$ to characterize the variability in the {\sl BP} and {\sl RP} bands.
These quantities describe the cross-epoch variance among the individual magnitude determinations that were used in DR2. They reflect the combination of magnitude {\sl measurement uncertainties} and of the intrinsic variability. For 
$G<17.5$, the photon-noise contribution to $\Delta_{G}$ is $\lesssim 0.03$~mag, as can be demonstrated from the plethora of non-varying sources. Substantially larger values $\Delta_{G}$ must then reflect intrinsic variability. This variance is typically determined from $\gtrsim 20$ epochs across 18 months; for most Cepheids it may therefore reflect a good approximation to the {\it rms} lightcurve variability.  For randomly sampled sinusoidal lightcurves 
$\Delta_{G}$ is 1/4 of the peak-to-peak variation. So, our adopted selection of $\Delta_{G}\ge 0.06$, will select Cepheid
lightcurves with peak-to-peak amplitudes of $\gtrsim 0.24$~mag.

\begin{figure}[ht]
\begin{center}
\includegraphics[width=\columnwidth]{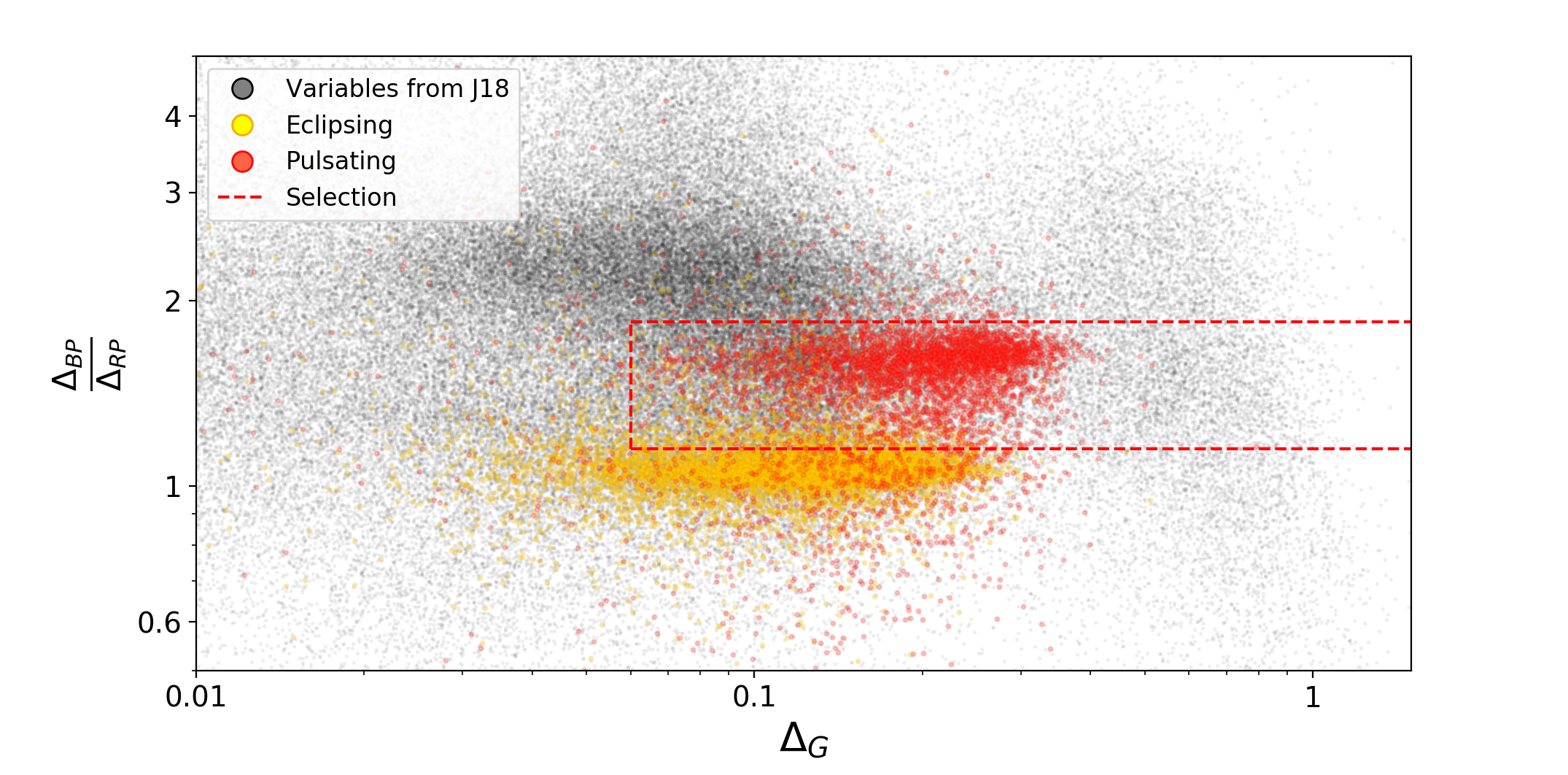}
\caption{
Selection of the Cepheid candidates by their Gaia variability and {\it color}-variability. The figure shows the G-band variability (see Eq.\ref{eqn:variability_definition}) vs the color-dependent variability $\frac{\Delta_{BP}}{\Delta_{RP}}$. This separates
quite neatly eclipsing binaries, pulsators, and cool, luminous giants, as indicated by the color-coding according to the classification provided by \citet{jayasinghe18}.
We use both $\Delta_{G}$ and $\frac{\Delta_{BP}}{\Delta_{RP}}$ in selecting Cepheid candidates, 
and specifically the range indicated by the dashed region. \label{fig:variability_selection}.
}
\end{center}
\end{figure}

It appears that $\Delta_{G}$ is a very effective way to select variable sources in Gaia.  But of course, such a crude \emph{rms}-measure cannot distinguish 
between lightcurves that vary because of pulsation, eclipses, secular changes, or other reasons. Among luminous stars, eclipsing binaries and long period variables (LPV) are the main contaminants.
But we can now exploit the fact that the effective temperatures of Cepheids, typically 5500~K - 6800~K, vary during the pulsation \citep{fukue15}; that means that the variations in a bluer (optical) pass-band should be higher. The quantity $\frac{\Delta_{BP}}{\Delta_{RP}}$ characterizes how much more sources vary in the blue \emph{vs.} the red. 
In Figure~\ref{fig:variability_selection} we show  $\frac{\Delta_{BP}}{\Delta_{RP}}$ for an extensive set of luminous variable objects. Figure~\ref{fig:variability_selection} shows therefore that the distribution of  $\frac{\Delta_{BP}}{\Delta_{RP}}$ is remarkably tri-modal, and that the three different groups (in $\frac{\Delta_{BP}}{\Delta_{RP}}$) map very well onto the independent classification by \citet[][hereinafter J18]{jayasinghe18}: 
eclipsing systems, where the flux variation arises from viewing geometry variations, show very little color dependence and $\frac{\Delta_{BP}}{\Delta_{RP}}\approx 1$; luminous, long-period variables (usually very red) near the top of the giant branch show distinctly stronger variations in BP. 
Bluer pulsating stars, including Cepheids, lie in between at $\frac{\Delta_{BP}}{\Delta_{RP}}\approx 1.4$. 
This shows that $\frac{\Delta_{BP}}{\Delta_{RP}}$ can be very effective in eliminating ``contaminants", and we adopt $1.15< \frac{\Delta_{BP}}{\Delta_{RP}}<1.85$ for candidate selection.

Finally, we select Cepheid candidates to be at low Galactic latitudes, $|b|<$20$^{\circ}$ (property ``4." above) and to have $G<17$; the latter selection criterion is motivated by the availability of useful ASAS-SN lightcurves for subsequent classification.

In addition, we apply an astrometric data quality, $\sqrt{ astrometric\_chi2\_al / ( astrometric\_n\_good\_obs\_al - 5)} < 2.5$, as recommended by Gaia DR2. Note that this is the only selection criterion that is not immediately related to a physical property of the objects themselves.
 
The Gaia DR2 SQL query encapsulating this selection function is listed in Appendix~A.


\subsection{Identification of Classical Cepheid from ASAS-SN lightcurves}
\label{sec:lightcurves}

The above selection criteria defined a pool of 29,737 Cepheid {\sl candidates}. 
As the initial selection neither assures that the lightcurve variations are periodic, 
nor that they have the shape expected for Cepheids, we may expect that the large majority of these candidates are ``contaminants''.
We then retrieved multi-epoch $V$-band photometry from the ASAS-SN automatic survey \citep{shappee14}, 
down to a limiting magnitude of $V\lesssim$18 mag, extracted as described in \citet{jayasinghe18} using image subtraction \citep{alard98} and aperture photometry on the subtracted images with a 2 pixel radius aperture. We corrected the zero point offsets between the different cameras as described in \citet{jayasinghe18}. The photometric errors were recalculated as described in \citet{jayasinghe19b}, and for individual epochs they range from 0.02 mag at $V$=11 mag to 0.16 mag at $V$=18 mag.

A fraction of $\lesssim 6$\% of candidates had to be excluded 
from subsequent analysis, because their lightcurves 
had less than 40 photometric measurements, excluding upper limits.

We thus analyzed the lightcurves of $\sim$30,000 $candidates$ to see which of them matched expectations 
for lightcurve shapes of classical Cepheids, first finding their periods then analyzing the shape of their folded lightcurves.
We started out with two independent period-finding analyses: 
a Lomb-Scargle \citep[LS,][]{Lomb,Scargle} periodogram, with lightcurves
folded by peak period of the power-spectrum; and we used a direct template-fitting approach to find the best period, described in \citet{inno15,kains19} (further details
on the fitting procedure, the code and some examples are given in Appendix~B and on the GitHub repository). The main difference between the LS approach and the template approach is that the former one assumes a generic sinusoidal shape for the lightcurve, while the latter uses  empirical lightcurves from observed Cepheids, and because it makes a stricter assumption on the lightcurve shape, 
can lead to wrong results for different kinds of variable sources.  
Thus, the template-fitting technique will give correct periods for lightcurve shapes similar to the ones of Cepheids, and incorrect results for other kind of variables or noisy lightcurves.

The difference between the most likely periods found with these 
two approaches was larger than 10\% for 49\% of the cases. In particular,
periods of 0.5,1,2 and 29.4 days were found for 20\% of the objects, presumably due to aliasing. 
Such a high fraction of discrepant period estimates need not be surprising or disconcerting, 
as we do not have direct {\it a priori} information whether the variability is even (short-term, $<1$~yr) periodic.
By removing all candidates with discrepant period estimates from the initial candidate list, 
we end up with a pool of $\sim$16,000 candidates with presumably periodic lightcurves,
and with typically 230$\pm$100 (standard deviation) ASAS-SN epochs.

Among them, classical Cepheids should form a distinctive family of light-curve shapes. 
We characterized these \emph{shapes} by fitting the lightcurves with
a 7$^{th}$-order Fourier series, 
assuming the best period from the template fitting method, 
which is expected to be more robust against possible outliers.

All the ASAS-SN folded lightcurves for the Cepheids we selected are
shown in the electronic version of Fig.~\ref{fig:atlas}, along with their Fourier-domain representation.
The results of such an analysis can be usefully quantified and illustrated by 
the ratio of the $2^{nd}$ and $3^{rd}$ Fourier amplitude to that of the first,
 $R_{21}$ and $R_{31}$, 
along with that the corresponding phase differences, $\phi_{21}$ and $\phi_{31}$. 

\begin{figure*}[ht]
\begin{center}
\includegraphics[width=0.95\textwidth]{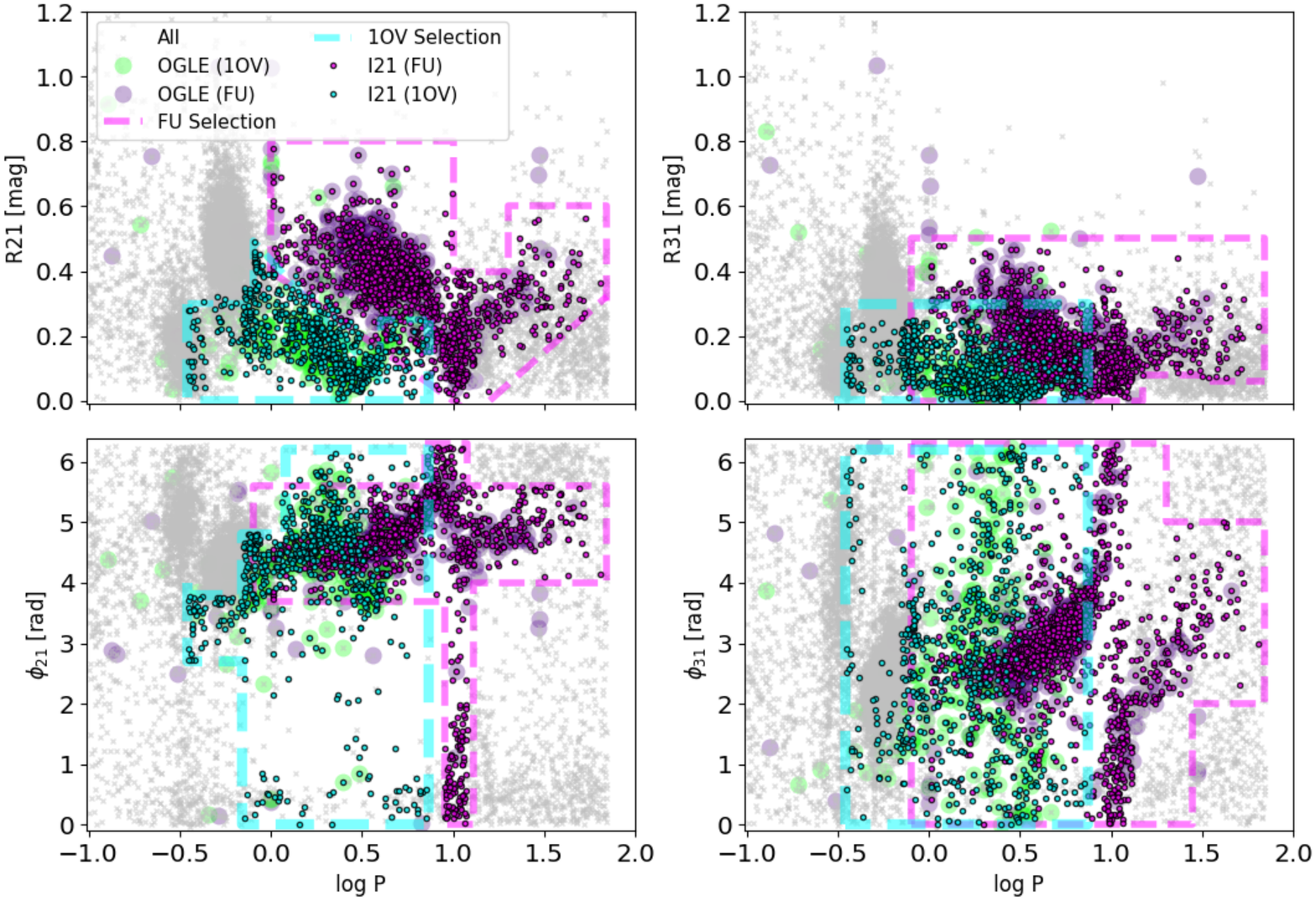}
\caption{Fourier-domain characterization of ASAS-SN photometry lightcurves 
for the candidate Cepheids, selected from Gaia DR2 (see Section~\ref{sec:DR2_selection}). 
The four panels show as a function of the pulsation period: the ratio of the $2^{nd}$ (top left) and $3^{rd}$ (top right) Fourier amplitude to the $1^{st}$ amplitude; and the corresponding phase shifts between these Fourier modes in the bottom panels.  
The dashed lines illustrate the Cepheid-selection criteria in this parameter space that we devised and adopted to select classical Cepheids pulsating in their fundamental mode (FU, magenta) and their first overtone mode (1OV, cyan). The Cepheids selected thereby are shown as cyan (FU), and magenta (1OV) dots; the open symbols indicate sources in common with the OGLE catalogs (U18,S20), as denoted through the inset label. 
Note that the distinct clumps of objects at periods shorter than 1 day are RR~Lyrae variables ($\sim$5,000 already known and $\sim$2,000 new objects), which are likely to be the more abundant contaminants in our selection for short-period Cepheids. 
\label{fig:sel1}
}
\end{center}
\end{figure*}

The distribution of the resulting Fourier parameters are 
shown in Fig.~\ref{fig:sel1} (grey crosses), and form a distinctly multi-modal distributions. 
Different parts of this distribution separate classical Cepheids from other variables, in particular from RR~Lyrae.
And among classical Cepheids these diagrams neatly separate 
those oscillating in the fundamental mode (FU) from those in the first overtone (1OV), 
as the pale green and lavender symbols classified by \citet{udalski19,sos20} 
from OGLE I-band lightcurves show. 

This diagram allows us now to identify potential FU Cepheids (dark purple) 
and 1OV Cepheids (dark green) among the totality of our candidates (grey dots).
We formalize this classification by the cuts shown in Fig.~\ref{fig:sel1} and listed in detail in Table~\ref{tab_sel}.

\begin{deluxetable*}{ll}
\tabletypesize{\scriptsize}
\tablecaption{Selection of Classical Cepheids in Lightcurve Shape Space
\label{tab_sel}}
\tablehead{
\colhead{Fundamental Pulsators}&
\colhead{First Overtone Pulsators}\\
}
\startdata
1.0$\le$P\tablenotemark{a}$\le$70   &   0.35$\le$P$\le$7.5 \\
(P$\ge$20 AND $R_{21}\le$0.6) OR ((10$\le$P$\le$20) AND ($R_{21}\le$0.4))  OR ((P$<$10) AND ($R_{21}\le$0.8)& (P$\ge$4 AND $R_{21}\le$ 0.25)  OR  (P$<$4  AND $R_{21}\le (0.45 -0.45\log P))$ \\
($R_{21}\ge (0.5\log P -0.6)$) AND ($R_{21}\ge (0.4 -0.4\log P$)) &   P$>$0.8 OR (P$\le$0.8 AND $R_{21}\le$0.3) \\
 $R_{31}\le$0.5 AND ((P$>$31 AND $R_{31}\ge$0.06) OR  P$\le$31) &$\phi_{21}\le$6.2\\
((15$\le$P$\le$30 )AND ($R_{31}\ge$0.08)) OR  P$>$30 OR P$<$15 &    P$>$0.7 OR    (P$\le$0.7 AND $\phi_{21}\le$3.8 AND $\phi_{21}>$2.7)\\
 ((P$\ge$30) AND ($R_{31}\ge$0.06)) OR P$<$30 &  P$>$1.2 OR    (P$\le$1.2 AND $\phi_{21}\le$4.8)    \\
 (P$\le$9 AND $\phi_{21}>$3.7) OR P$>$9 & \\
 ((P$\le$7 OR  P$\ge$12) AND $\phi_{21}<$5.6 ) OR 7$<$P$<$12 &\\
 (P$\ge$12 AND $\phi_{21}>$4.0) OR P $<$12 &\\
 (P$\le$3.5 AND $\phi_{31}<$3.5) OR P$>$3.5 &\\ 
 (P$\le$13 AND $\phi_{31}<$5.0 AND $\phi_{31}>$2.0) OR P$>$13 &\\ 
\enddata    

\tablenotetext{a}{Period measured in days}
\end{deluxetable*}

Remarkably, our well-defined, all-sky pre-selection of bright ($G<17$) candidates, 
followed by an algorithmic lightcurve classification results in a very large sample: 
a total of $\sim$1900 Galactic classical Cepheids, of which 1269 are FU and 622 are 1OV (note that we do not classify mixed mode pulsators). 
About 500 of them are new identifications (of which 25\% without any previous record). 
We provide the Gaia source\_id, the Gaia coordinates and all the parameters obtained from our analysis
for this sample in Table~\ref{tab:cat}, which is entirely available in the on-line version of the paper.

For the rest of the  paper, we refer to this sample as {\sl I21}.

\begin{deluxetable*}{lllllllllllll}

\tabletypesize{\scriptsize}
\tablecaption{Gaia DR2 ID and parameters from our analysis of all Cepheids in the I21 sample.
\label{tab:cat}}

\tablehead{
\colhead{Gaia DR2 source\_id}&
\colhead{P$_T$\tablenotemark{a}}&
\colhead{P$_F$\tablenotemark{b}}&
\colhead{N$_{epochs}$}&
\colhead{$<V>$}&
\colhead{$A_1$}&
\colhead{$R_{21}$}&
\colhead{$R_{31}$}&
\colhead{$\phi_{21}$}&
\colhead{$\phi_{31}$}&
\colhead{$\Delta_{G}$}&
\colhead{$\frac{\Delta_{BP}}{\Delta_{RP}}$}&
\colhead{Mode}
\\
\colhead{}&
\colhead{days}&
\colhead{days}&
\colhead{}&
\colhead{mag}&
\colhead{}&
\colhead{}&
\colhead{}&
\colhead{}&
\colhead{}&
\colhead{mag}&
\colhead{}&
\colhead{}
}
\startdata
175811260743708672	&	2.1136	&	2.114	&	126	&	13.805	&	0.309	&	0.507	&	0.332	&	4.375	&	2.615	&	0.247	&	1.555	&	F	\\
180464393952709504	&	1.08318	&	1.083	&	381	&	14.295	&	0.200	&	0.224	&	0.107	&	4.501	&	2.551	&	0.121	&	1.525	&	1O	\\
181620805307496192	&	1.66809	&	1.668	&	122	&	14.345	&	0.162	&	0.525	&	0.344	&	4.359	&	2.481	&	0.170	&	1.590	&	F	\\
186285139790902528	&	3.33083	&	3.331	&	316	&	12.921	&	0.141	&	0.078	&	0.039	&	3.594	&	1.25	&	0.068	&	1.532	&	1O	\\
186678245267045376	&	2.41531	&	2.415	&	153	&	14.828	&	0.174	&	0.524	&	0.277	&	4.186	&	2.478	&	0.129	&	1.615	&	F	\\
187324728037490048	&	4.91081	&	4.910	&	153	&	13.877	&	0.223	&	0.418	&	0.178	&	4.963	&	3.42	&	0.129	&	1.443	&	F	\\
188724234539584256	&	18.19102	&	18.21	&	256	&	10.515	&	0.366	&	0.165	&	0.122	&	4.942	&	2.124	&	0.185	&	1.489	&	F	\\
189726984845700480	&	5.2599	&	5.261	&	189	&	12.061	&	0.351	&	0.455	&	0.205	&	4.882	&	3.277	&	0.187	&	1.482	&	F	\\
189739251272296192	&	3.41275	&	3.414	&	189	&	12.478	&	0.360	&	0.475	&	0.266	&	4.579	&	2.758	&	0.147	&	1.444	&	F	\\
195001857519690112	&	1.82804	&	1.828	&	266	&	12.548	&	0.206	&	0.215	&	0.075	&	4.754	&	3.579	&	0.111	&	1.656	&	1O	\\
195373217573454464	&	3.85858	&	3.859	&	163	&	9.761	&	0.276	&	0.327	&	0.204	&	4.711	&	2.834	&	0.122	&	1.274	&	F	\\
197337185858157440	&	4.40485	&	4.405	&	267	&	12.123	&	0.340	&	0.389	&	0.17	&	4.613	&	2.75	&	0.222	&	1.756	&	F	\\
198681686717279616	&	2.65954	&	2.660	&	164	&	13.541	&	0.293	&	0.495	&	0.274	&	4.574	&	2.853	&	0.107	&	1.544	&	F	\\
200016111582079744	&	2.1206	&	2.121	&	400	&	13.803	&	0.391	&	0.479	&	0.278	&	4.363	&	2.543	&	0.252	&	1.381	&	F	\\
200044497021549056	&	1.87122	&	1.871	&	84	&	13.876	&	0.114	&	0.147	&	0.033	&	5.069	&	4.677	&	0.074	&	1.623	&	1O	\\
200256084290032256	&	2.1895	&	2.189	&	84	&	13.639	&	0.142	&	0.171	&	0.054	&	5.235	&	4.449	&	0.082	&	1.693	&	1O	\\
200708636406382720	&	11.62875	&	11.618	&	499	&	8.184	&	0.312	&	0.064	&	0.181	&	5.519	&	1.925	&	0.117	&	1.272	&	F	\\
201509768065410944	&	10.14831	&	10.148	&	103	&	9.572	&	0.268	&	0.191	&	0.129	&	4.257	&	6.095	&	0.131	&	1.820	&	F	\\
201574982848108416	&	10.28925	&	10.293	&	187	&	10.563	&	0.326	&	0.078	&	0.205	&	5.391	&	4.971	&	0.156	&	1.716	&	F	\\
204510644535613184	&	0.53381	&	0.534	&	84	&	13.281	&	0.123	&	0.243	&	0.148	&	3.377	&	1.078	&	0.081	&	1.603	&	1O	\\
204845480185040512	&	2.67975	&	2.680	&	243	&	15.338	&	0.104	&	0.087	&	0.009	&	5.874	&	3.079	&	0.063	&	1.507	&	1O	\\
206181318094220416	&	1.22802	&	1.228	&	227	&	12.452	&	0.138	&	0.180	&	0.025	&	4.356	&	2.56	&	0.074	&	1.637	&	1O	\\
206211859601169792	&	3.40703	&	3.407	&	147	&	13.141	&	0.374	&	0.451	&	0.254	&	4.608	&	2.818	&	0.215	&	1.515	&	F	\\
206577210999441536	&	13.84659	&	13.845	&	147	&	11.882	&	0.38	&	0.202	&	0.112	&	5.019	&	2.05	&	0.165	&	1.477	&	F	\\
207722317998325632	&	2.47689	&	2.477	&	142	&	13.199	&	0.188	&	0.365	&	0.141	&	4.424	&	2.592	&	0.124	&	1.654	&	F	\\
228673477705878400	&	2.31268	&	2.313	&	117	&	14.491	&	0.262	&	0.433	&	0.205	&	4.568	&	2.676	&	0.145	&	1.703	&	F	\\
228741784865956224	&	4.28995	&	4.291	&	117	&	11.173	&	0.31	&	0.427	&	0.223	&	4.737	&	3.219	&	0.207	&	1.681	&	F	\\
229602496313775872	&	2.0416	&	2.042	&	159	&	13.966	&	0.365	&	0.507	&	0.378	&	4.373	&	2.567	&	0.149	&	1.250	&	F	\\
232773865804463232	&	2.77515	&	2.775	&	159	&	13.232	&	0.175	&	0.047	&	0.027	&	5.585	&	4.622	&	0.073	&	1.515	&	1O	\\
247358200355952256	&	3.59323	&	3.594	&	109	&	14.234	&	0.099	&	0.402	&	0.28	&	4.617	&	2.995	&	0.154	&	1.632	&	F	\\
$\cdots$ &$\cdots$ & $\cdots$ & $\cdots$ &$\cdots$ & $\cdots$ & $\cdots$ &$\cdots$ & $\cdots$ & $\cdots$ &$\cdots$ & $\cdots$ & $\cdots$  \\
\enddata    
\tablenotetext{a}{Period obtained with the template fitting procedure }
\tablenotetext{b}{Period obtained with the LS method}
\end{deluxetable*}

\section{Completeness of the I21 Sample}
\label{sec:catalog_comp_1}
As mentioned in the introduction, we deem understanding the selection function, 
or ``completeness and purity'' of this I21 sample paramount. 

We have reason to believe that the purity of the sample is high ($\gtrsim 80\%$, see Section~\ref{sec:purity}), 
or that the contamination of the sample is relatively low, as the lightcurve shapes are of high discriminating value, 
and as the Gaia parallaxes are effective at eliminating low-luminosity (or more nearby) pulsators as contaminants; 
we will therefore focus first on completeness here.

Before detailing this aspect, it is worth taking a step back to spell out an operative definition of completeness. 
Most efforts to define ``a sample'' of astronomical sources set out to find objects within a well-defined set of criteria. 
Such samples should {\bf not} contain objects well outside of these criteria, and it would make little sense to attribute their absence 
to ``incompleteness''. 
E.g. the U18 sample contains no OGLE Cepheids 
in parts of the sky where OGLE did not observe. But this only implies that the global census of Cepheids in the Galaxy is incomplete, 
not that the U18 is highly incomplete\footnote{One may want to use the more general concept of a selection function, $S(\vec{p})$; this function, bound between 0 and 1, specifies the probability that an object with observable attributes $\vec{p}=\{position, magnitudes, colors, variability ...\}$ is in the sample. For a selection function it is understood that most of the (infinite) parameter domain has $S= 0$; also here the important part is to quantify $S(\vec{p})$ where it is not manifestly zero.}.

To explore the likely completeness of I21, we take the S19b as the comparison; ideally, one would like to have a gold standard sample
 with 100\% completeness and 0\% contamination, which is not practicable; but we take S19b as an approximation to that.

\begin{figure*}[ht]
\begin{center}
\includegraphics[width=0.80\textwidth]{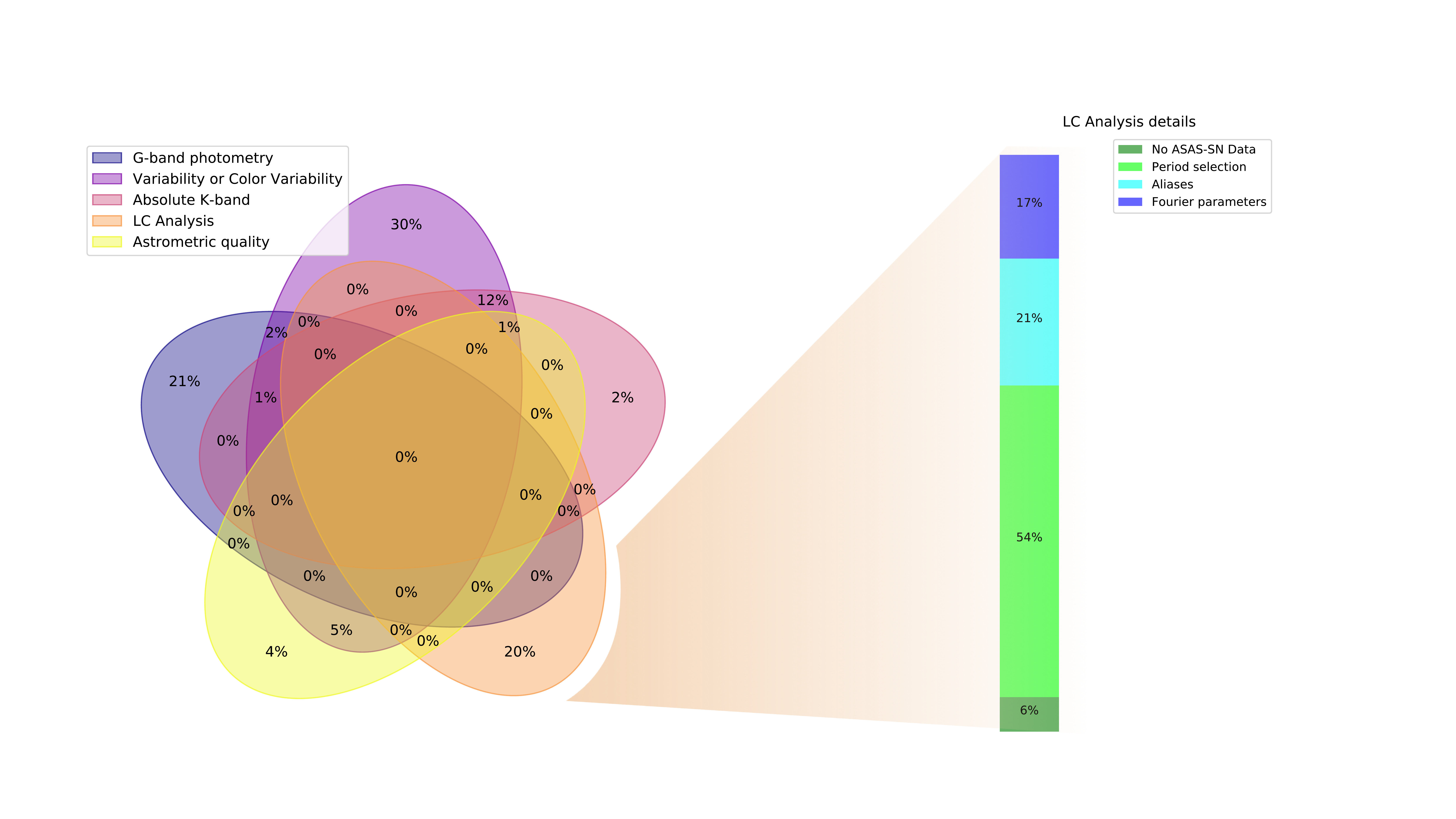}
\caption{Venn diagram for the Cepheids in the S19b catalog that were not selected by the two-step selection process of I21:
the initial Gaia DR2 query to obtain candidates, followed by the lightcurves analysis.
Among the criteria considered in GDR2 candidate selection are the G-band magnitude range (blue), variability amplitude (lilac), the K-band absolute magnitude (rose), and the astrometric data quality (yellow). The quality cuts and requirements in the lightcurve analysis are indicated in pale orange; the sub-aspects of the lightcurve analysis are specified in the right panel. There denote the reasons for not being included in I21: 
no photometry available from the ASAS-SN survey, different periods found with the two period-finding methods or period found beyond the chosen limits (0.3-70 days), manifest evidence of period aliasing, and failing to pass our cuts for the lightcurve shape Fourier analysis.
As discussed in Section~\ref{sec:catalog_comp_1},  ``completeness'' must be in reference to the 'objects' we set out to identify.
Therefore, objects in S19b that are outside the candidate search parameters of I21, do not contribute to I21s incompleteness.
What contributes, are objects missing because of data quality aspects: the dominant factor is the lightcurve analysis (20\%, with ambiguous periods being the main factor), with the only astrometric quality flags, as the only other factor, contributing to $\lesssim$5\%.
This implies a lower limit I21 sample completeness of $\sim$80\%, with higher completeness if the S19b sample has any contamination.\label{fig:venn_diagram}}
\end{center}
\end{figure*}

We illustrate this comparison with a Venn diagram \citep{Venn}, shown in Figure~\ref{fig:venn_diagram}, which provides a structured 
way of answering: how many of the Cepheids in S19b are {\bf not} in I21, and why?


At first glance this comparison may seem sobering with respect to completeness: of the 2483 objects in S19b only 1112 are also in I21, 
and 1371 of them seem to be missing. But Table~\ref{tab2} and Fig.~\ref{fig:venn_diagram} explains why that is: 1054 of the 1371 are not in I21, because they do not satisfy one or more of I21's initial selection criteria: the G-band magnitude range (blue), variability amplitude (lilac), the K-band absolute magnitude (rose), and the astrometric data quality (yellow), with the colors referring to the areas of the Venn diagram Fig~\ref{fig:venn_diagram}. 
So, only $\lesssim$60\% of these S19b Cepheids have intrinsic astrophysical properties that I21 were looking for. 
However, 317 of the S19b Cepheids are not in I21, even though their physical properties place them within I21's selection criteria. 
Almost all of them are missing, because they were excised during the lightcurve analysis (see Section~\ref{sec:lightcurves}); 
they were eliminated for a number of reasons, detailed in the right panel of Fig.~\ref{fig:venn_diagram}, 
with inconsistent period estimates among our two approaches as the leading reason.

These missing 317 Cepheids constitute true incompleteness. 
If we were to presume that the S19b is perfectly complete and pure, this implies a completeness 
for the I21 sample of $\sim$78\% (1112 of 1429 objects). If the S19b sample has any contamination the completeness may increase slightly.

In the subsequent Section (Sec.\ref{sec:catalog_purity_1}) we will address instead the question of the purity by comparing the I21 sample with both S19b and the 
pre-existent classification made on the basis of the same time-series data, i.e. the ASAS-SN catalog of variable stars \citep{jayasinghe18,jayasinghe19}.

\section{Purity of the I21 Sample}
\label{sec:catalog_purity_1}

Since we used photometric time-series data from the ASAS-SN survey, it is only natural to
compare our new classification with the one provided by the survey
(and available at \url{asas-sn.osu.edu/variables}) for the objects in common. 
The overlap between the two selections include only 71\% of our new sample,
namely 1354 objects, of which 55\% have been classified also as classical Cepheids by \citet{jayasinghe18,jayasinghe19}.
This leaves us with 610 candidate Cepheids in the I21 with a different classification in the ASAS-SN database. 
However, among those, 314 are indeed already classified as Cepheids by S19b or by \citet{sos20}, and therefore we will not discuss
further this subset, on the assumption that the OGLE sample does not include (or includes little) contamination. 

We therefore end up with only 296 objects with a previous different classification, 
which would lead to a contamination of $\sim$15\%. 
However, we can determine that purity of the I21 sample is actually significantly higher.

In fact, we find that the pulsation period reported in the ASAS-SN database 
is different for 36\% of the objects in common(for half of which the ASAS-SN 
period being double the period we found), and 
a wrongful determination of the period can easily lead to a miss-classification. 
By looking at the folded lightcurves, we verified that the new period 
is indeed the correct one for all the cases. However, we find that the objects classified
as "NON-PERIODIC" in ASAS-SN, show noisy lightcurves and therefore we consider them
as possibly misclassified objects (21). 
For the remaining ones (including the 188  with the same period),
we measure the skewness and kurtosis of their lightcurve shapes, in order to quantify 
how fast is the rising branch with respect to the decreasing branch, 
and then visually inspected all lightcurves individually.

By examining both this parameters's distribution and the lightcurves by eye, 
we were able to confirm our new classification for 78\% of the cases, 
with contaminants being mostly Type 2 Cepheids and Eclipsing binaries 
(the latter could be potentially removed by a clean cut on skewness $\lesssim$0.25).

Therefore, the contaminants amounts to 60 objects and hence the purity 
of the overlapping sample is 
$\gtrsim$90\%,
and the relative completeness
$within$ our selection criteria is 90\%. 


Finally, we compare the I21 overlap with the S19b sample and with the ASAS-SN sample
to assess their relative purity (933 objects). 
We find that 33\% of the objects in common between 
the three samples are classified as Cepheids by I21 and S19b but as different variables by ASAS-SN.
We also searched the OGLE data base for a match with the 
 641 objects in I21 with an ASAS-SN different classification: among the 167 matches,  54 objects were classified differently also in OGLE. Instead, by comparing the objects classified as Cepheids in both the ASAS-SN and I21 with the OGLE entire data base, we found  only 10 objects classified differently.

 

\begin{deluxetable*}{lrr}
\tabletypesize{\scriptsize}
\tablewidth{0pt}

\tablecaption{Relative completeness at each selection step. 
}
\label{tab2}
	
\tablehead{
\colhead{}&
\colhead{S19b sample (S19b)}&
\colhead{ASAS-SN sample (J18,J19)}
}
\startdata
All Cepheids with Gaia-DR2 entries \& within $|b|\le$20$^{\circ}$: & 2485 & 1035\\
G-band Magnitude:  6$\le G<$17 & 2056 & 1033\\
Absolute K-band Magnitude: $\varpi + \delta\varpi < 10^{0.2 (10. - \tilde{m}_K - M_K^{lim})}$ & 1907 & 1023\\
Variability: 0.06 mag $<\Delta_{G}<$2.5 mag \& 1.15$< \frac{\Delta_{BP}}{\Delta_{RP}}<1.85$ & 1607 & 852\\
Astrometric quality: $\sqrt( astrometric\_chi2\_al / ( astrometric\_n\_good\_obs\_al - 5)) < 2.5$ &1811 & 990\\
All Gaia-DR2 cuts:& 1429 & 827\\
ASAS-SN photometric time-series analyzed:  & 1410 & 824\\
Period from LS and template fitting within 10\%:&  1212 & 815\\
Fourier parameters within selected ranges in Tab.~\ref{tab_sel}:&  1112 & 743\\
&&\\
Relative completeness of I21:& 79\% & 90\%
\enddata    

\end{deluxetable*}

In the next Section, we address the purity and completeness 
of the I21 sample into the context of all Galactic Cepheid samples, 
many of which have some extensive overlap with I21. 
Nonetheless, a remarkable 126 Cepheids in our I21 sample are without prior published record, 
as we will show in the next discussion.

\section{The I21 Sample in Context: Comparison with Published Samples} 
\label{sec:catalog_comp}

In this Section we put the I21 sample into the context 
of all previously published samples, and verify if the relative completeness and purity discussed above
remains consistent. 
The basic motivation behind this step is that, as mentioned before, there is not a gold standard 
for the classical Cepheids' sample in the Milky Way to date, 
and while we believe the S19b can be adopted as a good approximation, 
we want to check if the completeness that we estimated for the I21 sample ($\sim$78\%) 
would either decrease or increase when considering all the public samples available.
Also, we want to determine how many of the Cepheids we identified are truly  "new" discoveries. 

We address the first objective in Sections~\ref{sec:compl_ref} and \ref{sec:purity}, and the second one in Section~\ref{sec:new}, but first we need to compile 
a master catalog from the available public databases to be used as the reference catalog, as we do in the following Section.

\subsection{The reference catalog of Galactic classical Cepheids, compiled from the literature}

In order to include all the classical Cepheids (or published candidates) currently known in the Milky Way,
we assembled a master catalog from the following public databases:

\begin{itemize}
      \item \textbf{OGLE}: The OGLE Collection of Galactic Cepheids, as published by \citet{udalski19}, 
      which includes 1,488 detection from OGLE-III and OGLE-IV (U18), plus 300 recently added by \cite[][,hereinafter S20]{sos20}, plus  the additional 1,142 objects collected by \citet{skowron19a,skowron19b} from the following catalogs: GCVS (702), ATLAS (168), ASAS (106), ASAS-SN (115), WISE (9), the VSX
      \footnote{www.aavso.org/vsx/index.php?view=search.top} (7) and specific papers: \citet{matsunaga16,tanioka17};
    \item \textbf{Gaia}: The Gaia Catalog of classical Cepheids, 
    	  as recently re-classified by R19, which includes 684 stars (520 FU, 147 1OV, 17 Multi-mode);
   \item \textbf{ASASSN}: The ASAS-SN catalog of classical Cepheids 
    cross-matched to Gaia DR2, with includes 2126 members, of which {\bf 1044} within $|b|\le$20$^{\circ}$  (812 FU and 232 1OV); 
     \item \textbf{Simbad}: $\sim$ 635 objects classified as $deltaCep$ in the SIMBAD database;
      \item \textbf{WISE}: $\sim$2100 Cepheids, including 779 new candidates identified from the WISE time-series
      data by Chen19, all assumed Fundamental mode pulsators. 
       \item \textbf{ZTF}: $\sim$1262 Cepheids, including 451 new candidates identified from the ZTF Survey by \citet[][hereinafter Chen20]{chen20}Chen20, all assumed Fundamental mode pulsators.
           \item \textbf{VSX}: $\sim$238 star out of the 3143 classified as DCEP, DEPCS, DCEP:, DCEP(B), DCEPS(B) in the VSX database within   20 degrees from the Galactic Plane but not already included in the previous catalogs;
\end{itemize}

\begin{figure*}[ht]
\begin{center}
\includegraphics[width=0.9\textwidth]{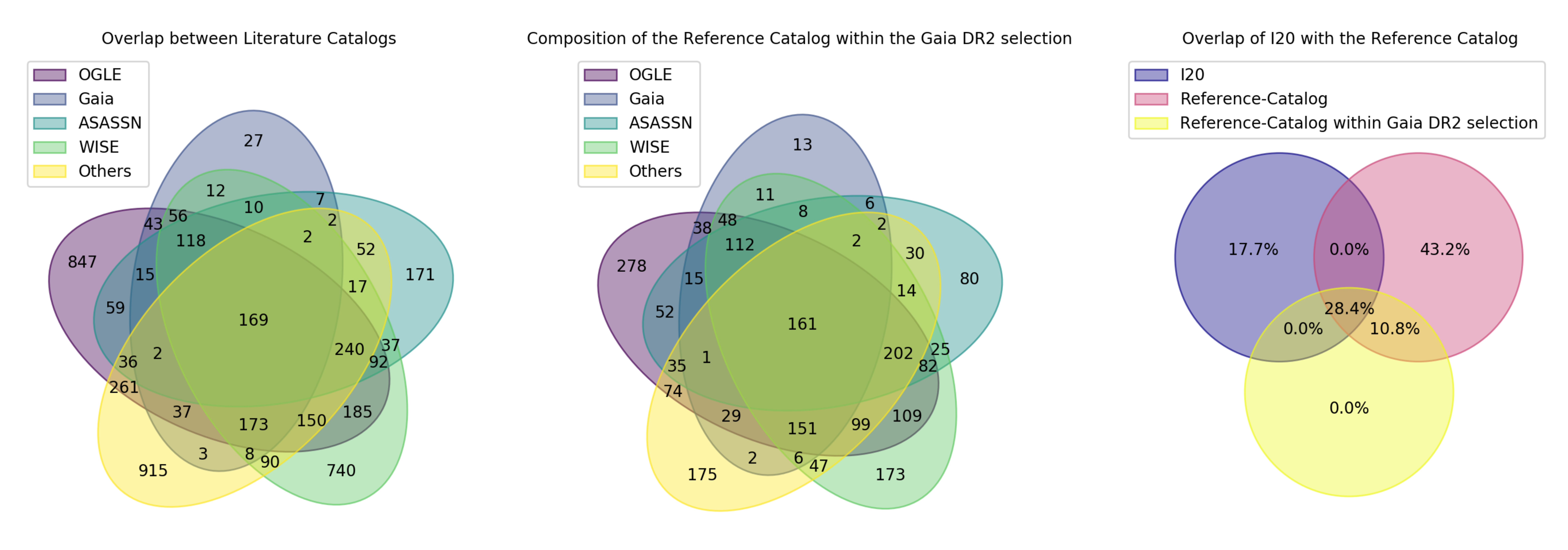}
\caption{Venn diagram to illustrate the overlap between the different literature samples 
used to build our reference catalog (left), the intersection between those catalogs 
and our initial Gaia DR2 query (middle), 
and the overlap between the latter and I21 (right). 
The pink petals of the Venn diagram  on the right shows 
that our Gaia-DR2 selection includes $\sim$50\%
of the currently known Cepheids, and its intersection with the yellow petal
 shows that number of Cepheids in the reference catalogs 
 $within$ our selection criteria {\bf but} not in the I21 is 12.7\%.
 Therefore, the relative I21 completeness is $\gtrsim$87\%.
\label{fig:venn_diagram_catalogs}}

\end{center}
\end{figure*}

The membership among almost all these catalogs has some, often substantial, overlap. 
But there is no published sample that mostly encompasses any of the other samples listed.  
This is a somewhat bewildering situation, which we try to illustrate in a Venn diagram,
shown in Figure~\ref{fig:venn_diagram_catalogs}. This Figure shows that there are e.g. $\gtrsim 1100$
Cepheids that are only in the OGLE sample, $\gtrsim 1000$ that are only in the WISE sample, $\gtrsim 700$ that are only in the ASAS-SN sample and $\gtrsim 1000$  that are only in the SIMBAD, VSX or ZTF samples (labeled as "Others"). 
It is beyond the scope of the present paper to sort out and discuss comprehensively the different overlap fractions in these catalogs.

But Fig.~\ref{fig:venn_diagram_catalogs} can serve to illustrate an elementary but central point in the present context: 
the ``completeness'' (or ``purity'') can only be sensibly discussed when the selection criteria are clearly stated; 
and then completeness should refer to the fraction of objects
included, whose properties fall {\sl within the stated selection criteria}. 

For example: a sample based purely on OGLE data will obviously be 100\% {\sl incomplete} 
in parts of the sky not covered by OGLE. 
Similarly, a Gaia based Cepheid catalog will obviously be 100\% incomplete in objects too reddened to be visible in the optical; any magnitude-limited catalog will be - by construction - fully incomplete in objects of fainter magnitudes, etc.. 
{\sl Therefore `completeness' is only a useful concept when phrased (implicitly or explicitly) 
as `complete within the stated selection criteria'. It is in this sense, that we now put our I21 sample into context.}

Thus, we build a comprehensive compilation
of Cepheids (or candidates) known in the Milky Way, by removing all duplicated sources, 
and adopting the pulsation period and mean luminosities (when available) by  
giving precedence to the catalog order in the above list.
Moreover, we also crossmatch the catalog with the OGLE database
in order to remove possible contaminants 
(54 objects with a different classification: 30 in J18, 6 in R19, 9 in Chen19 and 9 in literature).
This leads to a final literature compilation that includes 
4147 classical Cepheids (or published candidates) in the Milky Way (see Table\ref{tab1}), 
which we will now take as a reference catalog to check the completeness of the
new I21 sample, as done for the S19b and ASAS-SN catalogs in the previous section.

In fact, the I21 sample contains $\sim1900$ presumed classical Cepheids, of which 
$\sim 502$ are in no other Cepheids catalog (see right panel of Fig.~\ref{fig:venn_diagram_catalogs}). 
This leaves about 1390 objects overlapping with other catalogs, 
which is only $\lesssim 40$\% of the  $\sim 2500$ objects encompassed by those other catalogs. 
This could in principle be a consequence of I21 incompleteness, or of the contamination of the other catalogs. 

\begin{figure}[ht]
\begin{center}
\includegraphics[width=0.4\textwidth]{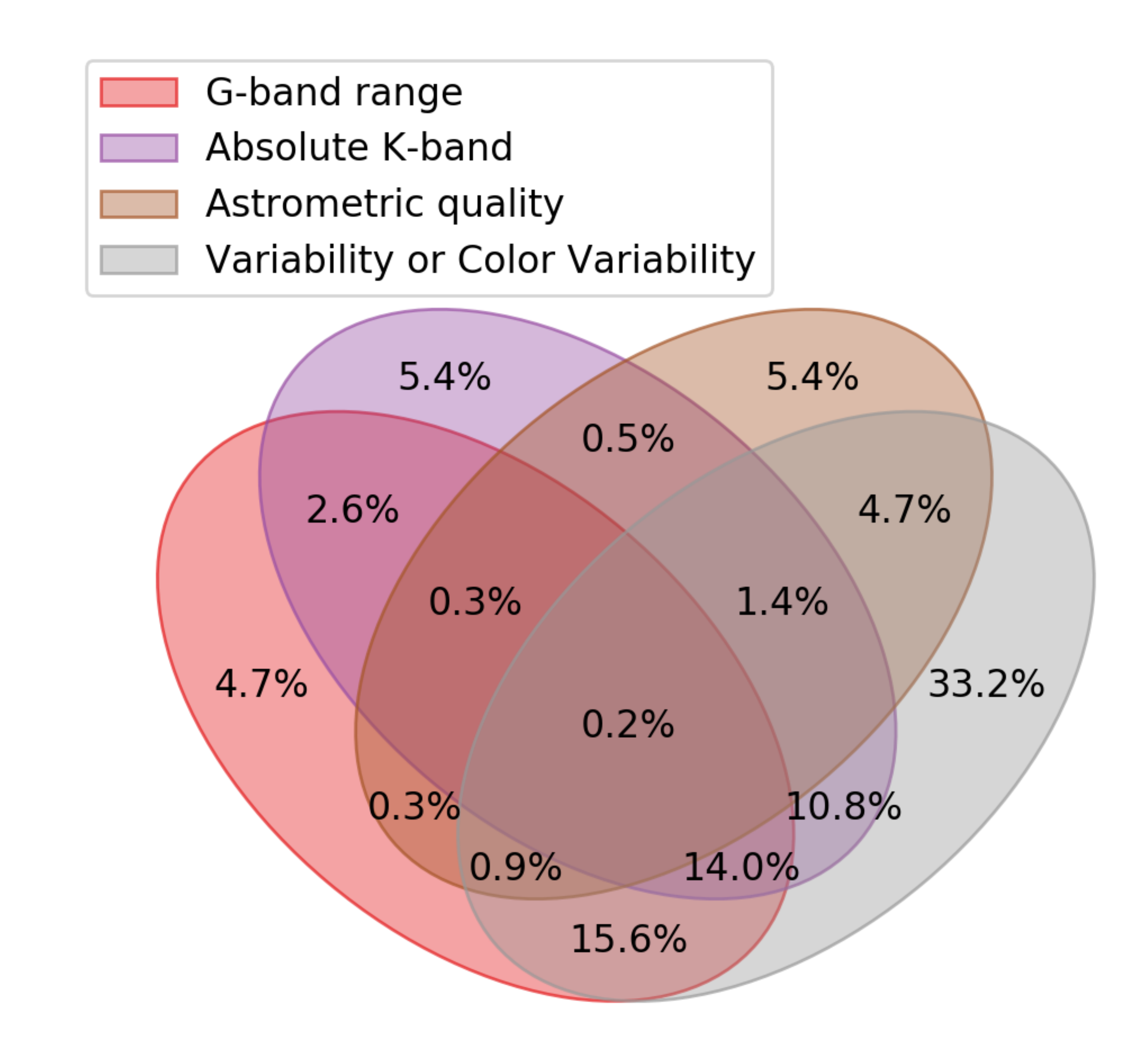}
\caption{Venn diagram for the Cepheids in the Literature catalog that are not in the I21 sample.\label{fig:venn_diagram_2}}
\end{center}
\end{figure}

In Figure~\ref{fig:venn_diagram_2} we show the Venn diagram 
of the objects contained in other catalogs, but not in I21: 
this could be because objects failed to pass varying combinations 
of the initial selection or lightcurve classification steps in I21. 
Fig.~\ref{fig:venn_diagram_2} illustrates that not passing the G-band magnitude limit, not varying enough or not being luminous enough are the most common reasons. 
However, I21 also used some {\sl data quality} cuts, i.e. criteria that do not refer to observables proprieties 
of the target objects, in particular the astrometric data quality. 
But, Fig.~\ref{fig:venn_diagram_2} illustrates that this eliminates only a small fraction of
objects ($\lesssim$5\%). This implies that the I21 sample is highly complete ($\sim$90\%)
{\sl within its stated target selection criteria},  

\begin{deluxetable*}{lcccccccr}
\tabletypesize{\scriptsize}
\tablewidth{0pt}

\tablecaption{Composition of the assembled reference catalog of Cepheids, 
the I21 sample and the final merged sample, which incorporates both of them. 
\label{tab1}}
\tablehead{
\colhead{Mode}&
\colhead{OGLE}&
\colhead{Gaia}&
\colhead{ASASSN}&
\colhead{Lit\tablenotemark{a}}&
\colhead{WISE}&
\colhead{ZTF}&
\colhead{Gaia-ASAS-SN (I21)}&
\colhead{Total\tablenotemark{b}}
}
\startdata
&\multicolumn{7}{c}{Initial Data-Set\tablenotemark{c}} \\
FU &  1840  & 520 & 812 & 971 & 2103 & 1238\tablenotemark{d} & 1269 & 3280 \\
1OV &  706  & 147 & 232 &  251 & ... & ... & 622  &805 \\
MULTI & 239  & 17 &  ... &  61  & ... &...  &...   &249  \\
&\multicolumn{6}{c}{Reference Catalog}\\
FU/1OV/MULTI &   &  &  &   & &  & & 4147\\
&\multicolumn{8}{c}{Merged Catalog including I21}\\
FU/1OV/MULTI &   &  &  &   & & & &   4721\\
\hline\hline
&\multicolumn{8}{c}{Subset within Gaia-DR2 selection} \\
\hline
FU/1OV/MULTI &1479&605&806&499&1269&651&1891& 2504\\
&\multicolumn{8}{c}{Overlap with I21} \\
FU/1OV/MULTI & 1131 & 503 & 744 & 410  &965 & 479& 1784 & 1373\\
\enddata    

\tablenotetext{a}{Includes VSX-cCs and SIMBAD-cCs}
\tablenotetext{b}{Without repetitions}
\tablenotetext{c}{Initial catalogs with original classification (duplicated objects included)}
\tablenotetext{d}{We assume fundamental-mode pulsation for the Cepheids in Chen+20}
\end{deluxetable*}


\subsection{Completeness of I21 with respect to the reference catalog}
\label{sec:compl_ref}
\begin{figure*}[!ht]
\begin{center}
\includegraphics[width=0.93\textwidth]{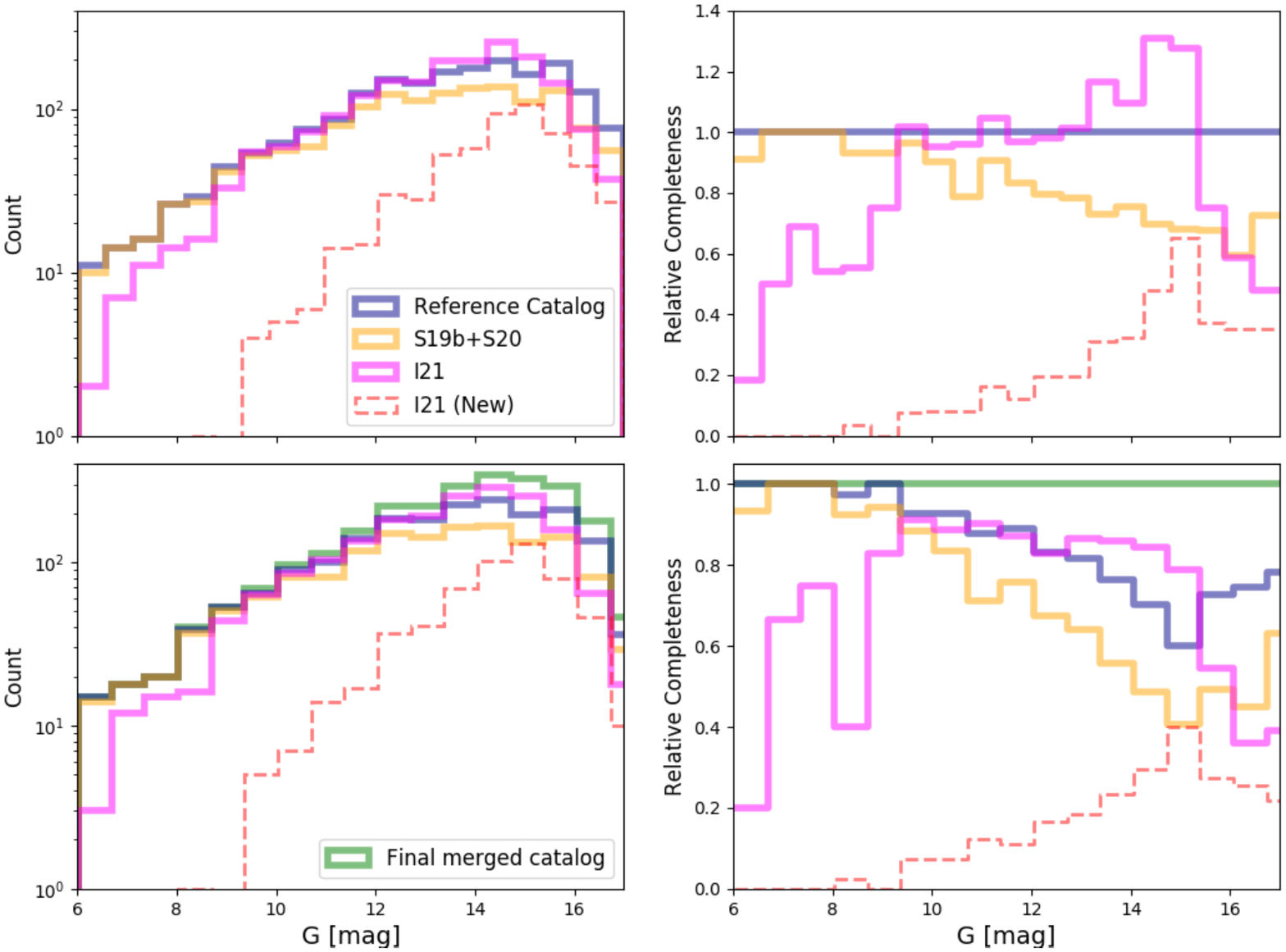}
\caption{Magnitude distribution (Left) and relative completeness (Right) 
of the different samples of Galactic Disk classical Cepheids 
\emph{within our Gaia DR2 selection parameters' range}.
In the top-left panel the histogram distribution of the 
Cepheids in the I21 sample (magenta line)
is compared to the one of the Catalog we assembled 
from the literature (which also include the catalog published by \citet{skowron19b}).
Note that we also plot the distribution of the newly identified Cepheids, 
which are not included in the literature-assembled catalog (red line) - thus also including the re-classified objects.
The ratio between the number of Cepheids in each luminosity bin 
is shown in the top-right panel for the same samples.
In the bottom-left panel, we show the same distribution from the the top panel, 
but we also include the one of the 
final merged catalog (i.e. the literature catalog and the new ones) in green. 
The completeness in the left-bottom panel is then computed 
on the basis of this new sample, and it shows that the Gaia-ASAS-SN 
sample we selected is largely complete for $G\gtrsim$9~mag and $G\lesssim$15~mag.
\label{fig:compl1}
}
\end{center}
\end{figure*}

In order to better characterize the completeness of the sample in terms of the
different stellar parameters included in our selection (luminosity, color, variability etc.),
we can study their distribution with respect to the reference catalog.
In Figure~\ref{fig:compl1}, we show the distribution
of the I21 sample as a function of the Gaia $G$-band magnitude 
and its relative completeness (right panel) 
compared to the catalogs mentioned above (top row, see labels).
By looking at this Figure, we can see that the I21 sample
is indeed more complete than the reference catalog (purple line)
between 10 and 17 $G$-mag, while the S19b sample alone is highly complete (90-100\%) 
in the range brighter than 8~mag, and complete to $\sim$70\% for fainter Cepheids.

If we now add the newly-identified Gaia-ASAS-SN Cepheids
to the reference catalog, creating a final merged catalog of about 
4721 Cepheids, of which 2504 within the Gaia DR2 parameters' range 
that we are exploiting (green line in Figure~\ref{fig:compl1}), 
the completeness of the reference catalog
drops by $\sim$20-40\% in the magnitude range between 12 and 17 mag,
whilst the I21 is indeed the most complete.


\subsection{Purity of the I21 sample by comparison with the Galactic reference catalog}
\label{sec:purity}

As mentioned above, there is a substantial overlap of the I21 
sample and previously known catalogs. In this Section, we
describe the overlap in detail, in order to provide quantitative 
estimates of our selection purity. 

Among the sources we selected, there are 
\begin{itemize}
\item {\bf 352
already included in the U18+S20 catalog, and we found the same period for 99\% of them.
For 14 Cepheids we identify a different pulsation mode, and, by visually inspecting the lightcurves,
we found that our classification seems to be correct for about 50\% of the cases.}
\item  {\bf 592} 
in the Gaia Catalog of Variable Stars, and we found the same period for 99\%
but a different classification for 15\% of them.

\item  {\bf 1354} 
objects in the classification by \citet{jayasinghe18,jayasinghe19}, of which 744 also classified as classical Cepheids.
Since we have already extensively discussed the overlap between these two samples in Section~\ref{sec:catalog_purity_1},
we will only remind the reader here that, on the basis of the lightcurve analysis for the
objects in common, we estimated a much lower contamination, around the 7\%.
We also found an additional 1070 objects classified as Cepheids by us and by other literature catalogs, but with a different classification in the ASAS-SN catalog. 
\item  {\bf 965}  
in the catalog of about 1,400 Cepheid 
candidates identified in the WISE data by \citet{chen19}.
However, the pulsation period we find is different for 14\% of the Cepheids 
and we find a different pulsation mode for 13\% of them.
Moreover, we find that among the 1209 objects in common between the 
Chen+19 sample and the OGLE sample, 13\% have a different pulsation period and
13\% are also classified
with a different pulsation mode by OGLE.
\item  {\bf 583} objects overlap with the ZTF sample by Chen20, and  82\% of them are also classified 
as classical Cepheids, while 104 objects are classified differently (among which 24 with a different period).

\item  {\bf 1358}  objects with a record in the SIMBAD database, of which  27\% classified 
as non-Cepheids. However, 20\% of the objects in the S19b catalog are 
also classified as non-Cepheids in SIMBAD, meaning that the SIMBAD 
classification can be out of date with respect to more modern studies 
of the same variables, and it should not be taken into account when estimating the purity of the sample.
\end{itemize}

In Table\ref{tab:cat2} we provide, for each Cepheid in I21, the name, data source, classification and classification reference from the literature for the objects that are also classified as Cepheids, while for the objects with also a different classification we provide the  classification from the ASAS-SN, OGLE or ZTF catalogs. 

\subsection{New Cepheids in I21}
\label{sec:new}

By removing all objects in common with other catalogs of Cepheids, we end up with 502 
potentially new Cepheids. 
Among those, 253 have been classified as different pulsating stars by J18,  99 (including 53 in common with J18) by OGLE (we already discussed these objects in Section~\ref{sec:catalog_purity_1}) and 79 (of which 47 in common with the previous two) by Chen20. 
Therefore, 502 are potential new Cepheids: 331 with a previous different classification, 126 new discoveries, 
and 45 with a record either in SIMBAD or VSX database (of which 4 as Cepheids) \footnote{Please note that these numbers can change over time, as new entries can be added to the ASAS-SN, SIMBAD or VSX database at anytime.}.
In order to validate the new discoveries and new classifications, we visually inspected the lightcurves 
and confirmed our classification for 65\% of them, with most of the unconfirmed classification being either eclipsing binaries and RR~Lyrae or lightcurves too noisy for providing a good classification. 
Therefore, the final number of contaminants we were able to identify is 9\%, 
confirming an overall purity higher than 90\%.

\begin{figure*}[!ht]
\begin{center}
\includegraphics[width=0.93\textwidth]{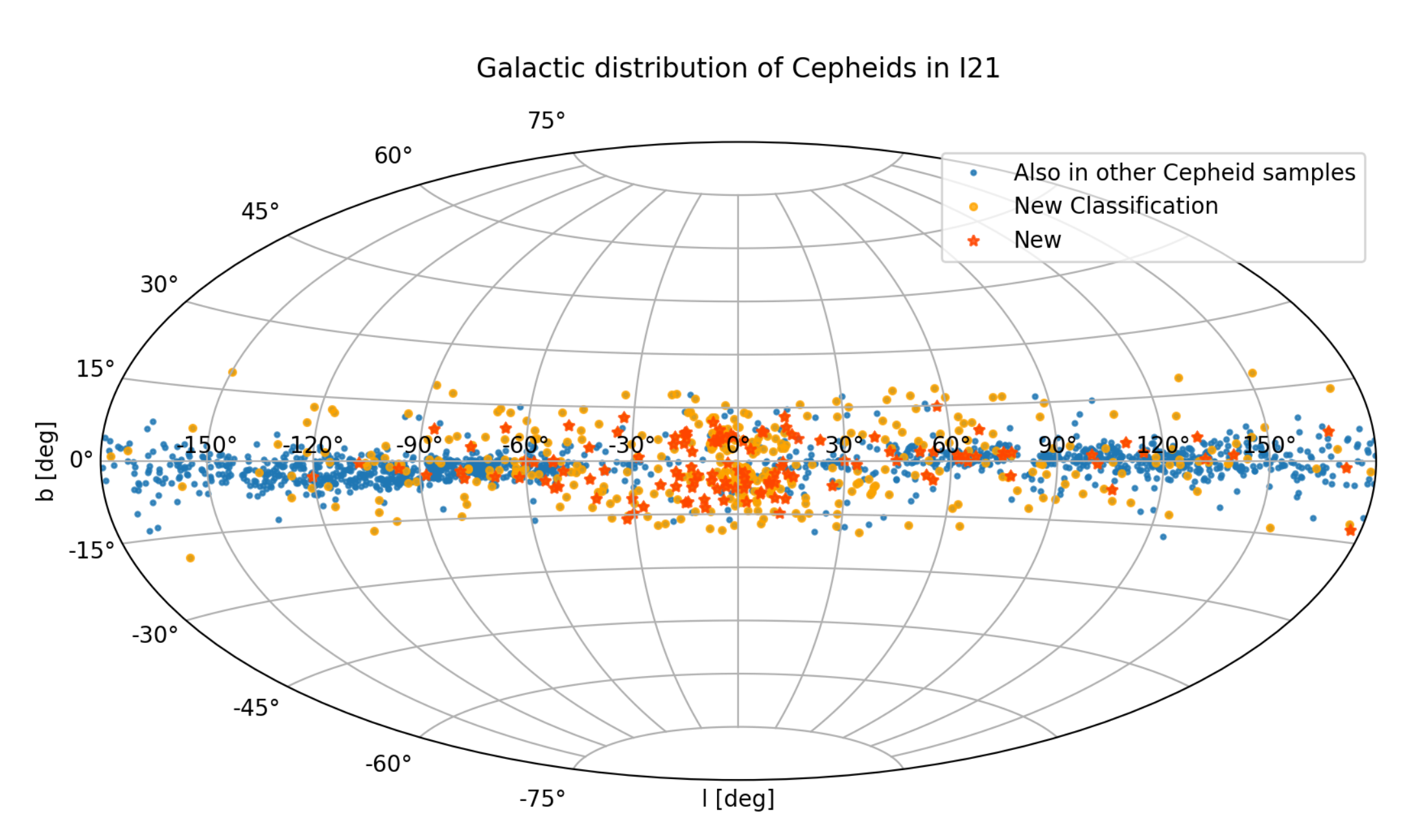}
\caption{Distribution of the I21 sample. The objects in common with other Cepheid samples are shown in blue, while the objects that have been re-classified as Cepheids in this work are indicated with orange dots. The 126 Cepheids for which there is no previous record in literature are shown as red stars.\label{fig:gal}
}
\end{center}
\end{figure*}

In summary, the Gaia-ASAS-SN catalog includes a sample of 1891 Cepheids that is complete and pure up to 90\% (9~mag$\lesssim G <$17~mag), of which 126 do not have any previous detection in the aforementioned catalogs, and  
331 previously classified into different variability types. The distribution in galactic coordinates of this sample is shown in Figure~\ref{fig:gal}.


\clearpage

\begin{turnpage}
\tabcolsep=0.11cm 
\begin{deluxetable}{lllllllllllll}
\tabletypesize{\scriptsize}
\tablewidth{0pt}
\tablecaption{Classification for the I21 Cepheids from other catalogs (the extended version of the table is available in electronic form).
\label{tab:cat2}}
\tablehead{
\colhead{Source\_id}&
\colhead{source\tablenotemark{a}}&
\colhead{star\tablenotemark{b}}&
\colhead{mode}&
\colhead{P}&
\colhead{Ref\tablenotemark{c}}&
\colhead{ASAS-SN Other Name\tablenotemark{d}}&
\colhead{ASAS-SN P}&
\colhead{ASAS-SN Other Type}&
\colhead{OGLE Other Name\tablenotemark{e}}&
\colhead{OGLE Other Type}&
\colhead{$\cdots$}
}
\startdata
153236088001040384 &  &  &  &  &  &  &  &  &  &  &  &\\
  175811260743708672 & VSX & NSVS6794216 & F & 2.1134 & S19b &  &  &  &  &  &  &\\
  180464393952709504 & ASAS-SN & J051348.97+311429.6 & 1O & 1.0832 & S19b &  &  &  &  &  &  &\\
  181620805307496192 & ATLAS & J076.9816+33.0565 & F & 1.6681 & S19b &  &  &  &  &  &  &\\
  183814674602852480 &  &  &  &  &  &  &  &  &  &  &  &\\
  186285139790902528 & ATLAS & J073.9358+36.3437 & 1O & 3.3305 & S19b &  &  &  &  &  &  &\\
  186678245267045376 & ASAS-SN & J050149.92+371627.8 & F & 2.4154 & S19b & J050149.92+371627.8 & 2.41563 & CWB &  &  &  &\\
  187324728037490048 & GCVS & IN~Aur & F & 4.9107 & S19b &  &  &  &  &  &  &\\
  188724234539584256 & GCVS & YZ~Aur & F & 18.1945 & S19b & J051521.97+400440.9 & NON-P & L &  &  &  &\\
  189726984845700480 & GCVS & GV~Aur & F & 5.2599 & S19b &  &  &  &  &  &  &\\
  189739251272296192 & GCVS & V335~Aur & F & 3.4129 & S19b &  &  &  &  &  &  &\\
  193684406370808448 &  &  &  &  &  &  &  &  &  &  &  &\\
  194690936247990528 &  &  &  &  &  &  &  &  &  &  &  &\\
  195001857519690112 & ASAS-SN & J052240.06+414302.0 & 1O & 1.828 & S19b &  &  &  &  &  &  &\\
  195373217573454464 & GCVS & Y~~Aur & F & 3.8595 & S19b &  &  &  &  &  &  &\\
  197337185858157440 & GCVS & GT~Aur & F & 4.4051 & S19b &  &  &  &  &  &  &\\
  198681686717279616 & GCVS & EW~Aur & F & 2.6595 & S19b &  &  &  &  &  &  &\\
  200016111582079744 & GCVS & FF~Aur & F & 2.1206 & S19b &  &  &  &  &  &  &\\
  200044497021549056 & ATLAS & J074.0591+40.2602 & 1O & 1.8712 & S19b &  &  &  &  &  &  &\\
  200256084290032256 & ZTF & ZTFJ044914.01+404059.8 & U & 2.1893 & Chen+20 &  &  &  &  &  &  &\\
  200708636406382720 & GCVS & RX~Aur & F & 11.6258 & S19b &  &  &  &  &  &  &\\
  201509768065410944 & GCVS & SY~Aur & F & 10.1446 & S19b & J051239.18+424955.2 &  NON-P & VAR &  &  &  &\\
  201574982848108416 & GCVS & AN~Aur & F & 10.2888 & S19b &  J045941.55+405010.2 &  NON-P & VAR &  &  &  &\\
  203413602509201792 &  &  &  &  &  &  J045215.80+415047.6 & 0.6898208 & RRAB &  &  &  &\\
  204510644535613184 & \citet{khruslov16} & GSC02901-00089 & 1O2O & 0.5338 & S19b & J044523.88+425520.1 & 0.5338166 & RRAB &  &  &  &\\
  204845480185040512 & ZTF & ZTFJ044412.39+435746.2 & U & 2.6796 & Chen20 &  &  &  &  &  &  &\\
  206181318094220416 & ZTF & ZTFJ045523.14+441815.5 & U & 1.228 & Chen20 &  &  &  &  &  &  &\\
  206211859601169792 & GCVS & V470~Aur & F & 3.4072 & S19b &  &  &  &  &  &  &\\
  206577210999441536 & GCVS & CY~Aur & F & 13.8414 & S19b &  &  &  &  &  &  &\\
  207722317998325632 & ASAS-SN & J052333.55+444536.5 & F & 2.4769 & S19b &  &  &  &  &  &  &\\
  228673477705878400 & CSS & J041331.0+411907 & F & 2.3127 & S19b &  &  &  &  &  &  &\\
  228741784865956224 & GCVS & SX~Per & F & 4.2901 & S19b &  &  &  &  &  &  &\\
  229602496313775872 & GCVS & GP~Per & F & 2.0416 & S19b &  &  &  &  &  &  &\\
  232773865804463232 & ATLAS & J065.6485+45.5819 & 1O & 2.7754 & S19b &  &  &  &  &  &  &\\
  236511827381118592 &  &  &  & &  &  &  &  &  &  &  &\\
  247358200355952256 & ATLAS & J062.4556+49.7355 & F & 3.5932 & S19b &  &  &  &  &  &  &\\
  248494232082353152 & GCVS & MM~Per & F & 4.1184 & S19b & J034523.42+480501.2 &  NON-P & VAR &  &  &  &\\
\enddata        
\tablenotetext{a}{Source of photometric data}
\tablenotetext{b}{Name of the star from literature sources}
\tablenotetext{c}{Reference of the catalog from which Period and Mode are adopted}
\tablenotetext{d}{Name for objects non classified as Cepheids in ASAS-SN}
\tablenotetext{d}{Name for objects non classified as Cepheids in OGLE}
\end{deluxetable}
\clearpage
\end{turnpage}

\section{Conclusion} 
\label{sec:conc}

We laid out and applied a straightforward approach for assembling a large ($\sim 1,900$), 
all-sky sample of classical Cepheids in the Milky Way, combining Gaia and ASAS-SN data. 
Our priority was to arrive at such a sample with a well-defined, reproducible and modellable 
selection function (rather than the most extensive compilation of such objects). 
Our sample, I21, is currently the only one that satisfies such selection-function quality criteria,
and has a completeness of $\sim90\%$ (9~mag$\gtrsim G<$17~mag).

A well-defined selection function is particularly important, if one wants to use Cepheids as tracers of the young star population across the Milky Way (e.g. Paper II, Inno et al. in prep). 
But it is also important, when selecting targets for ongoing and future spectroscopic surveys, such as SDSS-V \citep{Kollmeier2017} or 4MOST.  The selection parameters for this sample have been geared specifically towards SDSS-V, which has already started
taking data on this sample. In fact, SDSS-V aims to only observe samples of objects with algorithmically quantified selection function, making samples like this indispensable.

\acknowledgments
We are extremely grateful to C.S. Kochanek for his thoughtful reading of an early version of this manuscript. 
We thank the anonymous referee for the providing valuable comments.
LI acknowledges the "PON Ricerca e Innovazione: Attraction and International Mobility (AIM)"
program for support. 
This work was partially supported by the Sonderforschungsbereich 
SFB 881 (subproject A9) of the German Research Foundation (DFG). 
EP acknowledges support from the Centre national d'études spatiales (CNES).
This work has made use of data from the European Space Agency (ESA) 
mission Gaia (https://www.cosmos.esa.int/gaia), 
processed by the Gaia Data Processing and Analysis Consortium 
(DPAC, https://www.cosmos.esa.int/web/gaia/dpac/
consortium). Funding for the DPAC has been provided by national institutions, in
particular the institutions participating in the Gaia Multilateral Agreement.

\appendix
\section{Gaia query}

\begin{verbatim}
SELECT top 500000 *, sqrt(phot_g_n_obs)/phot_g_mean_flux_over_error as variability
FROM gaiadr2.gaia_source where phot_g_mean_mag < 17 
and
parallax + parallax_error <power(10.,(10.-  (phot_g_mean_mag -  1.72*(bp_rp) + 0.07*(bp_rp*bp_rp))-1.)/5.) 
and 
sqrt( astrometric_chi2_al / ( astrometric_n_good_obs_al - 5)) < 2.5
and 
sqrt(phot_g_n_obs)/phot_g_mean_flux_over_error > 0.06  
and 
(sqrt(phot_bp_n_obs)/phot_bp_mean_flux_over_error )/(sqrt(phot_rp_n_obs)/phot_rp_mean_flux_over_error) 
< 1.85 
and 
(sqrt(phot_bp_n_obs)/phot_bp_mean_flux_over_error )/(sqrt(phot_rp_n_obs)/phot_rp_mean_flux_over_error) 
> 1.15 
and 
abs(b) < 20
\end{verbatim}
\section{Template Fits and Power Spectrum}
\label{ap:c}
\subsection{Template Fitting method}
We used the python packages $gatspy$\footnote{https://zenodo.org/badge/DOI/10.5281/zenodo.14833.svg} in order to obtain the Lomb Scargle periodogram\citep{LS_multiband} of the lightcurves for each source, then we fold the lightcurve according to each of the 3 top periods found by the algorithm and finally we use $scipy$ to perform a fit of the folded lightcurve by means of an empirically-calibrated templates. The procedure then selects the period at which the template-fitting produces finds the lowest residuals (chi-squared minimization).
\subsection{LS-based method}
We used the python package $astropy$ \citep{astropy1,astropy2} in order to obtain a Lomb-Scargle periodogram (we adopted the Nyquist factor=300), and then identify the period corresponding to the frequency of the peak.

%
\begin{figure}[!ht]
\begin{center}
\includegraphics[scale=0.32]{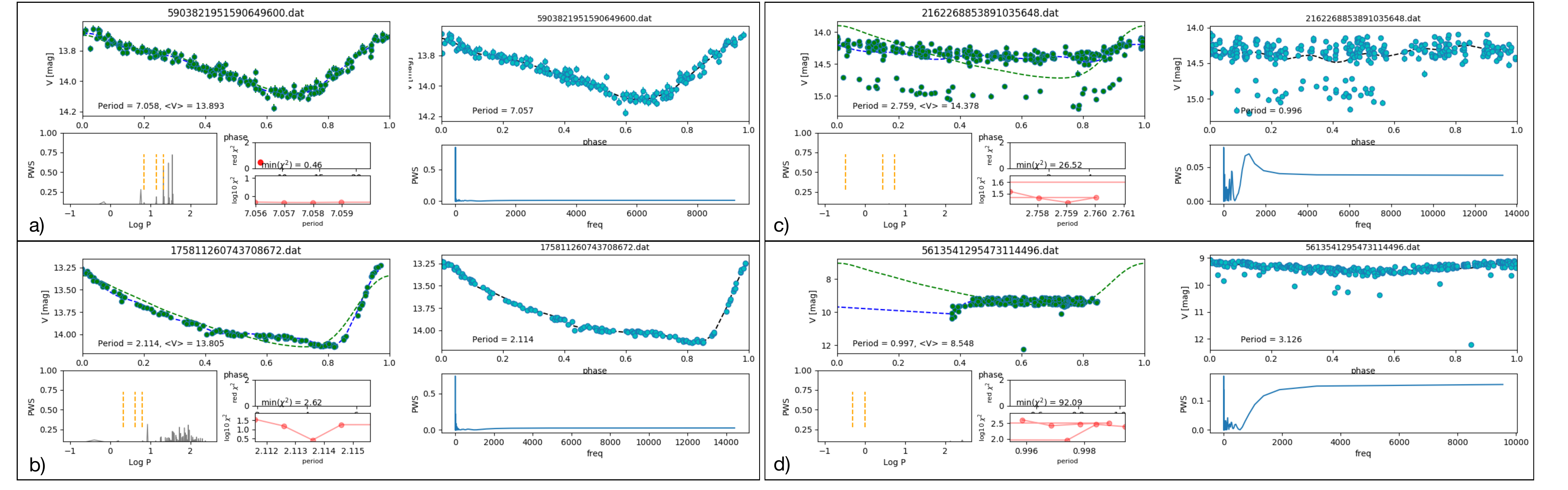}
\caption{Output of our light-curve fitting code for four different objects with Gaia-ID as labeled in the top of each plot. For each object in each panel ($a)$,$b)$,$c)$,$d)$) we show the output form the template-fitting procedure (left) and from the application of the LS periodogram (right). The lightcurves shown in the top sub-panel is folded according to best value of the period found by each procedure. In the case of the template-fitting, this is done by performing a chi-squared minimization of the data with the respect to the template curve (green dashed line) for values of the period shown by the orange line in the power-spectrum shown below. The LC folded accordingly, is then fitted with a Fourier series (blue-dashes line). The second method shown in the left is simply based on finding the period corresponding to the highest peak in the LS periodogram shown below (cyan line, frequency space). 
Case $a)$ and $b)$ show two examples of objects for which the two methods obtain the same value for the period. Case $c)$ shows an example where the template-fitting method find the correct period despite the presence of outliers, while the simple periodogram finds the aliasing around 1 day. Case $d)$ shows a case in which the template fitting finds the aliasing, while the LS methods finds the correct period.
\label{fig:fits}
}
\end{center}
\end{figure}

\section{Lightcurves atlas}
\begin{figure}[ht]
\begin{center}
\includegraphics[scale=0.42]{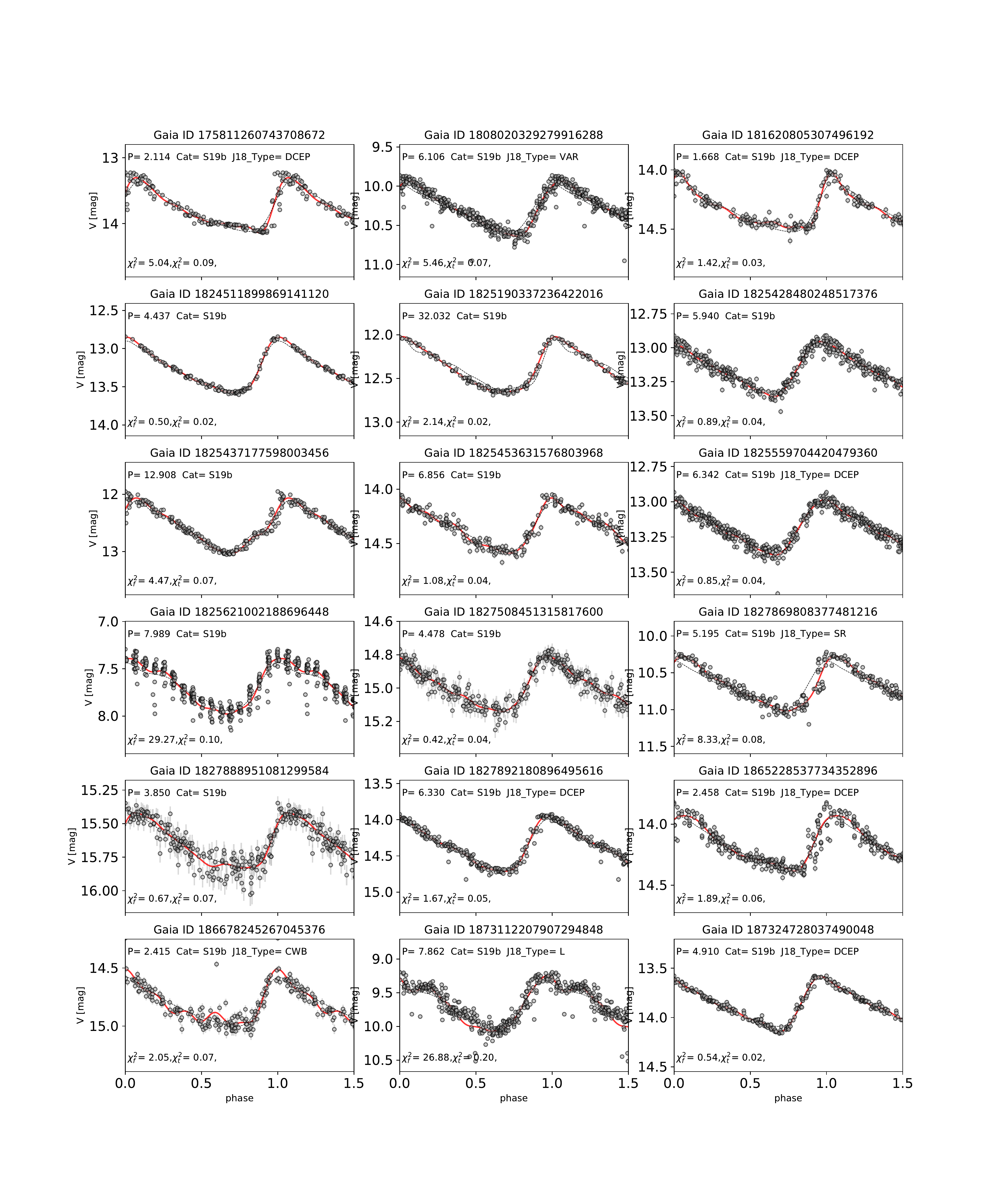}
\caption{Period-folded ASAS-SN lightcurves of Cepheids in our I21 sample. 
The red lines show the $7^{th}$-order Fourier fit, whose coefficients are used in the lightcurve classification (see Figure~\ref{fig:sel1}), while the thin black line is the best-fitted template. 
\label{fig:atlas}.
}
\end{center}
\end{figure}
\begin{figure}[ht]
\begin{center}
\includegraphics[scale=0.42]{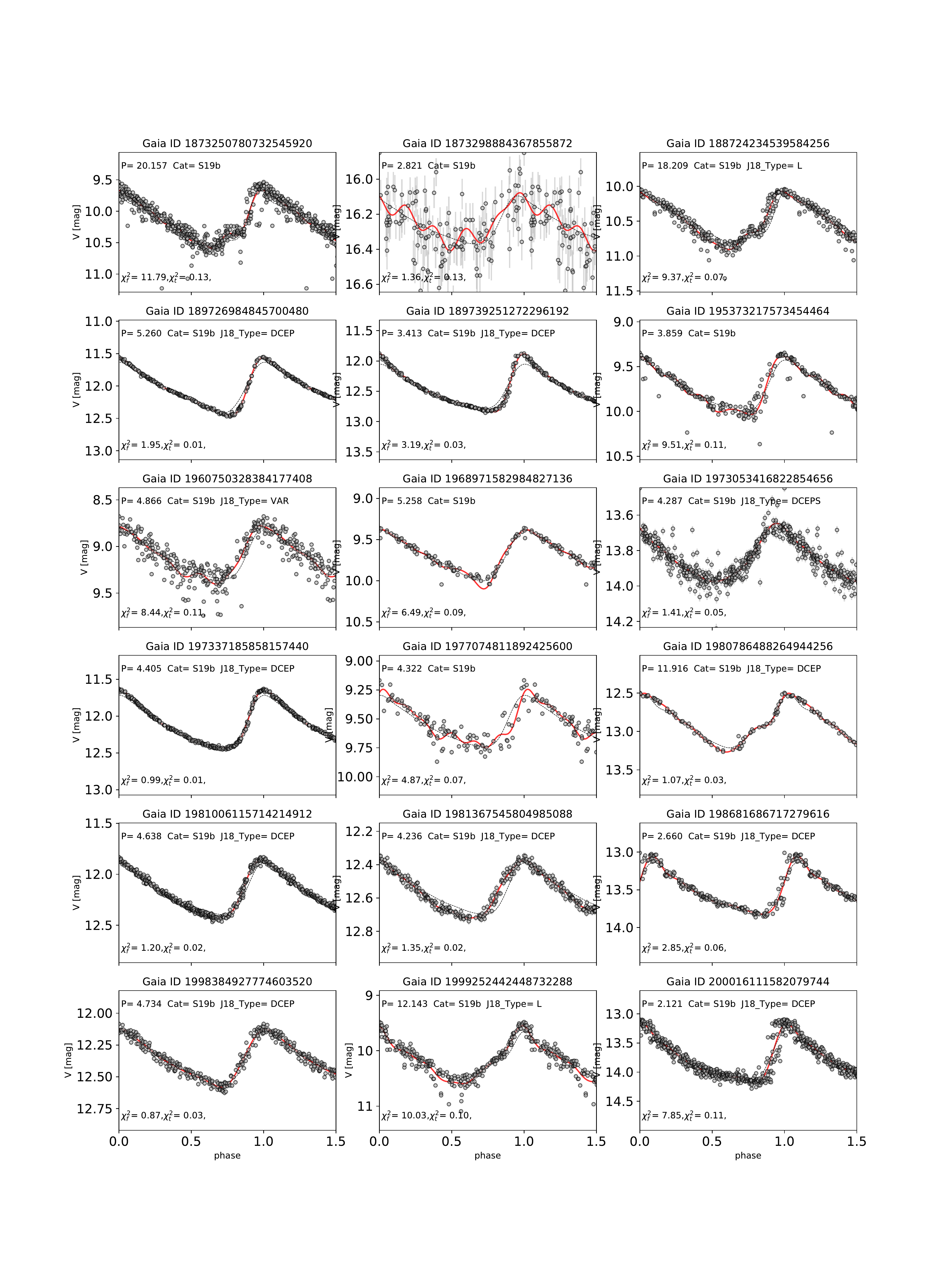}
\caption{cont.
}
\end{center}
\end{figure}

\begin{figure}[ht]
\begin{center}
\includegraphics[scale=0.42]{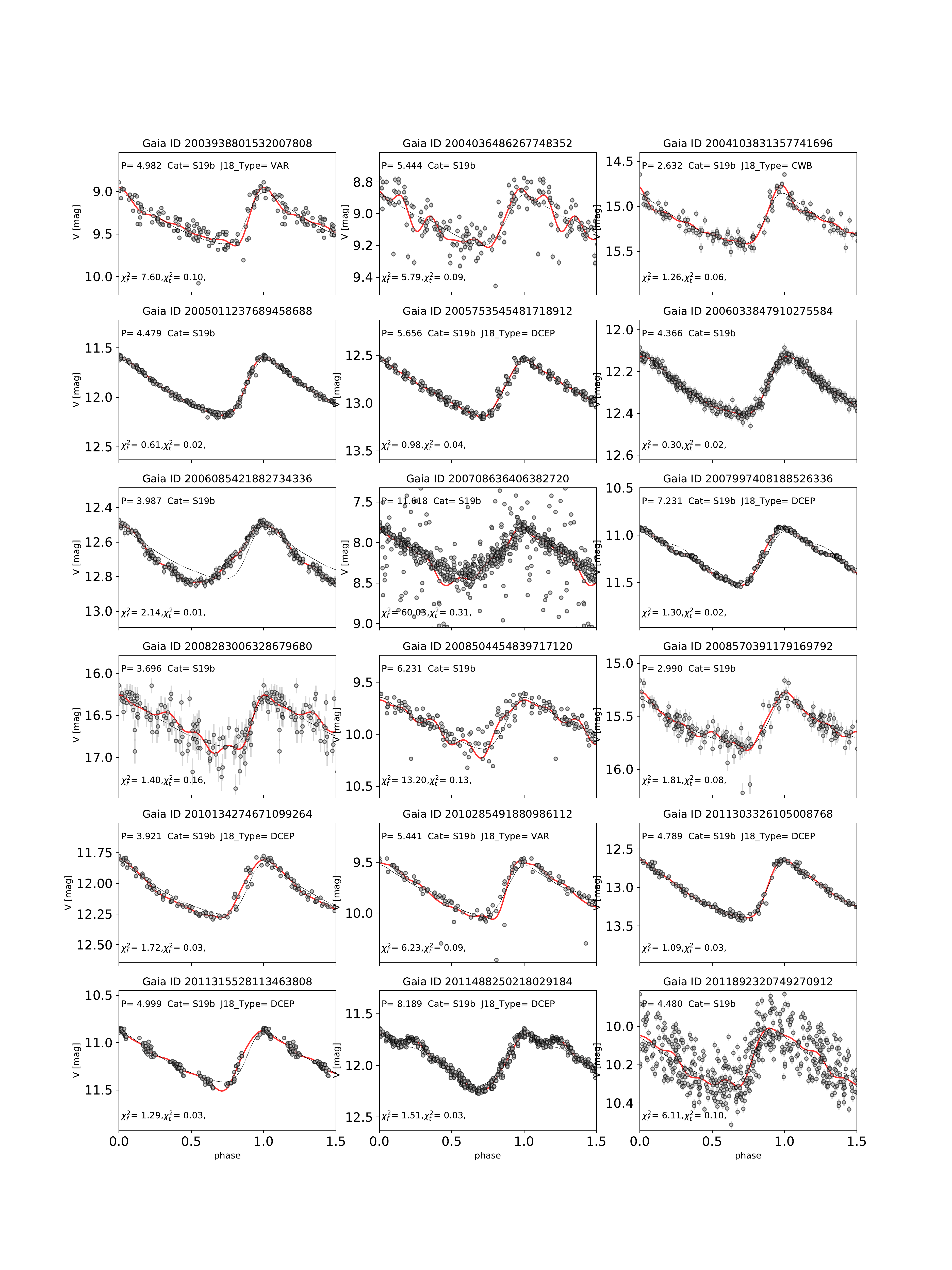}
\caption{cont.
}
\end{center}
\end{figure}

\end{document}